\documentclass{jetpl}
\twocolumn

\lat

\title{Microscopical derivation of Ginzburg-Landau-type functionals for alloys 
and their application to studies  of antiphase and interphase boundaries}

\rtitle{Ginzburg-Landau functionals for alloys with application to antiphase boundaries}

\sodtitle{Microscopical derivation of Ginzburg-Landau-type functionals for alloys 
and their application to studies of of antiphase and interphase boundaries}

\author{V.\,G.\,Vaks\/\thanks{e-mail: vaks@mbslab.kiae.ru}}

\rauthor{V.\,G.\,Vaks}

\sodauthor{Vaks}

\address{Russian Research Centre ``Kurchatov Institute", 123182 Moscow, Russia}

\dates{}%{}

\abstract{
The earlier-described cluster methods are used to generalize the Ginzburg-Landau 
gradient expansion for the free energy of an inhomogeneous alloy 
to the case of not small values of order parameters and
variations of composition. The results obtained reveale
a number of important differences  with the expressions used in the 
 phenomenological phase-field  approach.
Differential equations relating the local values of
concentration and order parameters within the antiphase or interphase boundary
(APB or IPB) are derived. These equations  are applied to study 
the segregation at APBs near the phase transition lines; the structure of 
  APBs and IPBs near  tricritical points; 
  wetting  APBs in phases with single and several order parameters; 
and also some effects of anisotropy of APBs under L1$_0$ and L1$_2$-type orderings.}

\PACS{05.70.Fh, 61.50.Ks}

\begin{document}

\maketitle

Studies of inhomogeneous alloys attract interest from both fundamental 
and applied points of view, in particular, in connection with microstructure evolution under
phase transformations [1\ch 12]. Typical inhomogeneities in such problems are 
antiphase  or interphase boundaries (APBs or IPBs) which separate
differently ordered domains or  different phases. Both experimental and
theoretical studies show that in situations of practical interest the APB or IPB width
usually much exceeds the interatomic distance [5\ch 12]. Therefore, Ginzburg-Landau (GL) type
gradient expansions can be used to describe the free energy of such
states even though the order parameters and concentration variations here
 are typically not small, contrary to  assumptions of the standard GL
 theory. Employing such generalized GL functionals (suggested first  by Cahn and Hilliard 
\cite{Cahn-Hilliard}) is now referred  to as the phase-field method,
and it  is widely used for most different systems, see e. g.  [6\ch 8]. 
However, a number of simplifying asumptions are usually employed in  this  phenomenological 
approach, and so its  relation  to more consistent theoretical treatments
remains unclear. Recently microscopical cluster methods have been developed
for inhomogeneous alloys [9\ch 13].
Below I use  these methods to derive the GL functionals
and then  apply them
for studies of APBs and IPBs.

To be definite, I consider a binary alloy A$_c$B$_{1-c}$ at $c\leq 1/2$. Various 
distributions of atoms over lattice sites $i$ are described by the mean
occupations $c_i=\langle n_{i}\rangle $ where  $n_{i}$
is unity when the site $i$ is occupied by  atom A and
 zero otherwise, while averaging is taken, generally, over the space- and
time-dependent distribution function \cite{BDPSV}. The free energy 
$F\{c_i\}$ in the cluster description 
can be written as a series \cite{VS}:
\begin{equation}
F=\sum_if^i=\sum_i\biggl(F_1^i+\sum_jF_2^{ij}+\cdots\sum_{j,\ldots k}F_m^ {ij\ldots k}\biggr).
\label{F}
\end{equation}
Here $f^i$ is the free energy per site $i$;  
$F_1^i=T[c_i\ln c_i+(1-c_i)\ln (1-c_i)$] is the mixing entropy contribution;
$F_l^ {i\ldots k}=F_l(c_i,\ldots c_k)$ is the contribution 
of interactions within $l$-site cluster of sites $i,\ldots k$; and $m$ is the maximum cluster size
considered. The simplest mean-field approximation (MFA) and the pair-cluster one 
(PCA) correspond to 
neglecting   many-site contributions $F_{m>2}$ in (\ref{F}), while
in a more refined, tetrahedron cluster approximation -- TCA (that should be used,
in particular, to adequately describe the L1$_2$ and L1$_0$-type orderings \cite{BDPSV})
Eq.  (\ref{F}) includes also 4-site terms $F_4^{ijkl}$   \cite{VS}.

For the homogeneous ordered structure, the mean occupation $c_j=c({\bf r}_j)$
at site $j$ with the lattice vector ${\bf r}_j$  can be written as a superposition of 
concentration waves with some superstructure vectors ${\bf k}_s$ 
[8\ch 12]:
\begin{equation}
c_j=c+\sum_s\eta_s\exp(i{\bf k}_s{\bf r}_j)\equiv
\sum_p\eta_p\exp(i{\bf k}_p{\bf r}_j).
\label{c_j}
\end{equation}
Here amplitudes $\eta_s$ can be considered as order parameters; the last
expression  includes also term with
 $\eta_p=c$ and  ${\bf k}_p=0$; and for simplicity both parameters 
$\eta_s$ and factors $\exp(i{\bf k}_s{\bf r}_i)$ are supposed to be real
which is the case, in particular, for the B2, L1$_0$ and  L1$_2$-type order.
For weakly inhomogeneous states, amplitudes $\eta_p$ in (\ref{c_j}) are not
constants but smooth functions of coordinates ${\bf r}_i$. Thus
functions $f^i\{c_j\}$ in (\ref{F}) can be expanded in powers of differences 
$\delta c_j=\sum_p\delta\eta_p^j\exp(i{\bf k}_p{\bf r}_j)$ where
\begin{equation}
\delta\eta_p^j=\eta_p^j-\eta_p^i={\bf r}_{ji}\nabla\eta_p+
\frac{1}{2}\sum_{j}r_{ji}^{\alpha} r_{ji}^{\beta}\nabla_{\alpha\beta}\eta_p.
\label{delta_eta}
\end{equation}
Here  $\nabla\eta_p=\partial\eta_p^i/\partial r_i^{\alpha}$, 
$\nabla_{\alpha\beta}\eta_p=\partial^2\eta_p^i/\partial r_i^{\alpha}\partial r_i^{\beta}$, 
${\bf r}_{ji}=({\bf r}_{j}-{\bf r}_{i})$, and the
summation over repeated Cartesian indices $\alpha,\beta =1,2,3$ is implied.
After substitution of these expressions  into
Eq. (\ref{F}) one can proceed from summation over $i$ to integration over continuous
 variable ${\bf r}={\bf r}_i$. Making also standard manipulations with part-by-part integration
of terms with $\nabla_{\alpha\beta}\eta_p$  \cite{Kikuchi-Cahn-62} one obtains
 for the GL functional:
\begin{equation}
F=\frac{1}{v_a}\int d^3r\Bigl[\sum_{p,q}g_{pq}^{\alpha\beta}
\nabla_{\alpha}\eta_p\nabla_{\beta}\eta_q+f\{\eta_p\}\Bigr].
\label{F_GL}
\end{equation}
Here $v_a$ is volume per atom; $f\{\eta_p\}$ is function $f^i\{c_j\}$ in (\ref{F}) 
 averaged over all sublattices with $c_j$  given by Eq. (\ref{c_j});
and  $g_{pq}^{\alpha\beta}$ is given by the expression:
\begin{eqnarray}
g_{pq}^{\alpha\beta}=-\frac{1}{2}\sum_jr_{ij}^{\alpha} r_{ij}^{\beta}
S_{ij}^i\exp[i({\bf k}_p{\bf r}_j-{\bf k}_q{\bf r}_i)]+\nonumber\\ 
+\frac{1}{2}\sum_{k,j;\, j\neq i}r_{ki}^{\alpha}(r_{ji}^{\beta}-r_{ki}^{\beta})
S_{kj}^i\exp[i({\bf k}_p{\bf r}_j-{\bf k}_q{\bf r}_i)]
\label{g_pq}
\end{eqnarray}
where $S_{kj}^i=\partial^2f^i/\partial c_k\partial c_j$. Note that the last term of
(\ref{g_pq}) is nonzero only when all three sites $i$, $j$ and $k$ are
different, and so it is present only when the non-pairwise contributions
$F_{m>2}^{ij\ldots k}$ in (\ref{F}) are taken into account, such as the TCA  terms $F_4^{ijkl}$
\cite{VS}.

For the L1$_2$ and L1$_0$ phases in FCC alloys, Eq. (\ref{c_j}) includes
waves with  three  vectors ${\bf k}_s$: ${\bf k}_1=(100)2\pi/a$, ${\bf k}_2=(010)2\pi/a$,
and ${\bf k}_3=(001)2\pi/a$ where $a$ is the lattice constant [8\ch 15]. 
The local order within APBs in these phases 
 can be  described
by the distribution of amplitudes of these waves $(\eta_1,\eta_2,\eta_3)$ of the type
$(\zeta ,\eta ,\eta )$  and corresponds to the tetragonal symmetry
 \cite{Braun-Cahn,BDPSV,Vaks-01}. 
To illustrate the form of terms $g_{pq}^{\alpha\beta}$ in (\ref{g_pq}), 
we present their MFA and PCA expressions for this type local order.
 Tensors $g_{pq}^{\alpha\beta}$ here can be described
in terms of their ``transverse" and
``anisotropic" components,  $g_{pq}^{\bot}=g_{pq}^{22}=g_{pq}^{33}$
and  $g_{pq}^a=(g_{pq}^{11}-g_{pq}^{22})$. 
Using for simplicity
the 2-neighbor-interaction model:
$v_{n>2}=0$, one obtains in the MFA:
\begin{eqnarray}
g_{\zeta\zeta}^{\bot}&=&-\frac{1}{2}a^2v_2;\,
g_{\zeta\zeta,\eta\eta}^{a}=\pm\frac{1}{2}a^2v_1;\,
g_{\eta\eta}^{\bot}=\frac{1}{2}a^2(v_1-2v_2);\nonumber\\ 
g_{cc}^{\bot}&=&-\frac{1}{2}a^2(v_1+v_2);\qquad 
g_{cc}^{a}=g_{\eta c,\zeta c,\eta\zeta}^{\bot,a}=0,
%=g_{\zeta c}^{\bot,a}=g_{\eta\zeta}^{\bot,a}=0
\label{g-L1_2-MFA}
\end{eqnarray}
while the PCA expressions for $g_{pq}^{\bot}$ and $g_{pq}^a$ are:
\begin{eqnarray}
g_{\zeta\zeta ,cc}^{\bot}&=&\frac{1}{4}a^2(-\varphi_1^+\pm\psi_1^+
-\chi_2^+-\varphi_2^{dd});\ \ g_{\zeta c}^a=\frac{1}{4}a^2\varphi_1^{-};\nonumber\\ 
g_{\zeta\zeta ,cc}^a&=&\frac{1}{4}a^2(\varphi_1^+\pm\psi_1^+);\ g_{\zeta c}^{\bot}
=\frac{1}{4}a^2(-\varphi_1^{-}-\chi_2^++\varphi_2^{dd});\nonumber\\ 
g_{\eta\eta}^{\bot}&=&\frac{1}{2}a^2(\varphi_1^{ab}-2\,\chi_2^+);\quad
g_{\eta\eta}^a=-\frac{1}{2}a^2\varphi_1^{ab};\nonumber\\ 
g_{\eta\zeta,\eta c}^{\bot}&=&\frac{1}{4}a^2(\pm\psi_1^{-}-2\varphi_2^-);\quad
g_{\eta\zeta,\eta c}^a=\pm\frac{1}{4}a^2\psi_1^{-}.
\label{g-L1_2-PCA}
\end{eqnarray}
Here plus or minus in $(\pm)$ corresponds to the first or the second pair of lower indices in
the left-hand-side of Eqs. (\ref{g-L1_2-MFA})-(\ref{g-L1_2-PCA});
%equations;%
\begin{eqnarray}
\varphi_n^{ij}&=&-Tf_n/R_n^{ij};\qquad \ \ \ 
\varphi_n^{\pm}=\frac{1}{2}(\varphi_n^{ab}\pm \varphi_n^{dd});\nonumber\\
\psi_n^{\pm}&=&\frac{1}{2}(\varphi_n^{ad}\pm \varphi_n^{bd});\qquad
\chi_n^{\pm}=\frac{1}{2}(\varphi_n^{aa}\pm \varphi_n^{bb});\nonumber\\
R_n^{ij}&=&[1+2f_n(c_i+c_j-2c_ic_j)+f_n^2(c_i-c_j)^2]^{1/2};
\label{R_n}
\end{eqnarray}
$f_n=\exp(-v_n/T)-1$ is the Mayer function; and index $i$ or $j$ 
 equal to $a$, $b$ or $d$
corresponds  to the mean occupation $c_i$ or $c_j$ of one of three
different sublattices:
\begin{equation}
c_a=c+\zeta+2\eta;\quad c_b=c+\zeta-2\eta;\quad c_d=c-\zeta.
\label{c_abc}
\end{equation}
If the TCA is used, the nearest-neighbor contributions $\varphi_1^{ij}$ in Eqs. 
(\ref{g-L1_2-PCA}) are replaced by the relevant tetrahedron contributions
$S_{ij}^{i}$ presented in Ref. \cite{VZK}. For the B2 order, 
there is only one order parameter $\eta_s=\eta$ \cite{BDPSV};
terms $g_{pq}^{\alpha\beta}$ have a cubic symmetry: 
   $g_{pq}^{\alpha\beta}=\delta_{\alpha\beta}g_{pq}$; 
and the MFA and PCA expressions for $g_{pq}$ are similar to those for $g_{pq}^{\bot}$ 
in Eqs. (\ref{g-L1_2-MFA}) and (\ref{g-L1_2-PCA}).

In the phase field method, terms $g_{pq}^{\alpha\beta}$ are  assumed to not
depend on local parameters $\eta_r$, and so
to be zero at $p\neq q$
 ``on considerations of symmetry" [6-8].
Eqs. (\ref{g-L1_2-MFA})-(\ref{g-L1_2-PCA}) show that it may correspond only 
to the simplest MFA, while in more accurate approaches, such as PCA and TCA,
the dependences $g_{pq}^{\alpha\beta}(\eta_r)$ can be significant.

Let us now consider
the case of a plane APB (or IPB) 
when parameters $\eta_p$ in (\ref{c_j}) depend only on the distance 
$\xi ={\bf rn}_0$  where
${\bf n}_0=(\cos\alpha,\sin\alpha\cos\varphi,\sin\alpha\sin\varphi)$ is  normal to the 
APB plane.
 To find the equilibrium structure, one 
should minimize the functional (\ref{F_GL}) over functions $\eta_p(\xi)$
at the fixed total number of atoms \cite{Kikuchi-Cahn-62}. 
Let us first consider the APB between two B2-ordered domains. 
Then variational equations for the order parameter $\eta(\xi)$ and the concentration $c(\xi)$
have the form:
\begin{eqnarray}
g_{\eta\eta}\eta''+g_{\eta c}c''&+&\frac{1}{2}g_{\eta}^{\eta\eta}(\eta')^2
+g_c^{\eta\eta}\eta'c'+\nonumber\\ 
&+&(g_c^{\eta c}-\frac{1}{2}g_{\eta}^{cc})(c')^2=\frac{1}{2}f_{\eta};\nonumber\\ 
g_{\eta c}\eta''+g_{cc}c''&+&(g_{\eta}^{\eta c}-\frac{1}{2}g_c^{\eta\eta})(\eta')^2
+g_{\eta}^{cc}\eta'c'+\nonumber\\
&+&\frac{1}{2}g_{c}^{cc} =\frac{1}{2}(f_c-\mu).
\label{eqs_eta-c}
\end{eqnarray}
Here prime means taking derivative over $\xi$; the lower index  $\eta$ or $c$ means 
taking derivative over $\eta$ or $c$;
and $\mu$ is the chemical potential.
 At $\xi\to\infty$
functions $c$ and $\eta$ tend to their equilibrium values, $c_0$ and $\eta_0(c_0)$.

Multiplying  first and second Eq. (\ref{eqs_eta-c}) by $\eta'$ and $c'$, respectively, 
summing them, and integrating the result, one obtains the first integral of this
system of equations:
\begin{equation}
g_{\eta\eta}(\eta')^2+2g_{\eta c}\eta'c'+g_{cc}(c')^2=\Omega (\eta,c)
\label{first-int}
\end{equation}
where $\Omega$ is the non-gradien part of the local excess grand canonical potential per atom:
\begin{equation}
\Omega  =f(\eta, c)-f_0-\mu (c-c_0),
\label{Omega}
\end{equation}
and index zero at the function means its value  at $\eta =\eta_0$ and $c=c_0$. 

For what follows it is convenient to consider the order parameter 
$\eta$ as an independent variable, while $c$ and $\Omega$,
 as its functions determined by Eqs. (\ref{eqs_eta-c})-(\ref{Omega}).
Then the dependence $\eta^{\prime}(\eta)$  is determined  by Eq. (\ref{first-int}):
\begin{equation}
\eta^{\prime}=d\eta/d\xi=(\Omega/G)^{1/2}
\label{eta_xi}
\end{equation}
where $G$ is $(g_{\eta\eta}+2g_{\eta c}\dot c+g_{cc}{\dot c}^2)$, and $\dot c$ is $dc/ d\eta$.
Using Eq. (\ref{eta_xi}) one can exclude $\eta^{\prime}$ from the system
of equations (\ref{eqs_eta-c}) and obtain the differential equation for
$c(\eta)$ to be called the composition-order 
equation (COE):
\begin{eqnarray}
[\ddot c\,(g_{\eta c}^2-g_{cc}g_{\eta\eta})&+&\Phi]\,2\,\Omega\, /G=\nonumber\\ 
=(\mu-f_c)(g_{\eta\eta}&+&g_{\eta c}\dot c)+f_{\eta}(g_{\eta c}+g_{cc}\dot c)
\label{COE}
\end{eqnarray}
where $\Phi$ is a linear  function of derivatives $g_{r}^{pq}$:
\begin{eqnarray}
\Phi =(g_{\eta c}&+&g_{cc}\dot c)\Bigl[\frac{1}{2}g_{\eta}^{\eta\eta}+g_c^{\eta\eta}\dot c
+\Bigl(g_c^{\eta c}-\frac{1}{2}g_{\eta}^{cc}\Bigr){\dot c}^2\Bigr]-\nonumber\\ 
-(g_{\eta\eta}&+&g_{\eta c}\dot c)
\Bigl(g_{\eta}^{\eta c}-\frac{1}{2}g_c^{\eta\eta}+g_{\eta}^{cc}\dot c
+\frac{1}{2}g_{c}^{cc}{\dot c}^2\Bigr).
\label{Phi}
\end{eqnarray}
Because of the equilibrium conditions:
$f_{\eta}^0=0$, $f_c^0=\mu $, function $\Omega (\eta )$ 
(\ref{Omega}) at $\eta\to\eta_0$ is proportional to
$(\eta -\eta_0)^2$. Therefore, the initial value $\dot c (\eta_0)$ can be found by
taking the  $\eta\to\eta_0$ limit  of Eq. (\ref{COE}).

For the given solution $c(\eta )$ of COE,
the coordinate dependence
$\eta (\xi)$ is determined  by integrating Eq. (\ref{eta_xi}):
\begin{equation}
 \xi =\xi_1+\int_{\eta_1}^{\eta}d\eta\,(G/\Omega )^{1/2}
\label{xi_eta}
\end{equation}
where the reference point $\xi_1$ is determined  by the choice of  value $\eta_1=\eta (\xi_1)$.
For the symmetrical APB for which
$\eta\to\pm\eta_0$ at $\xi\to\pm\infty$,
functions $c$, $\Omega$ and $G$ are even in $\eta$, and
it is natural to put  $\xi_1=0$ at $\eta_1=0$. But
for an IPB separating the ordered and the
disordered phase, values $\eta_1\to 0$ correspond to 
$\xi\to (-\infty )$, and so $\xi_1$ should be chosen at some
intermediate value $\eta_1$.

The surface energy $\sigma$ and the surface  segregation $\Gamma$ 
is the excess of the grand canonical potential
 and of B atoms, respectively,   per unit area 
\cite{Kikuchi-Cahn-62}. Taking into account Eqs. (\ref{first-int})-(\ref{eta_xi}) 
one obtains for the surface energy: 
\begin{equation}
\sigma =\frac{2}{v_a} \int_{\eta_{\rm min}}^{\eta_0}d\eta\,(G\,\Omega)^{1/2}
\label{sigma}
\end{equation}
where $\eta_{\rm min}$ is $(-\eta_0)$ for an APB and zero for an IPB, while the
surface segregation is given by the expression:
\begin{equation}
\Gamma=\frac{1}{v_a}  \int_{-\eta_0}^{\eta_0}d\eta\, (c_0-c)\,(G/\Omega)^{1/2}.
\label{Gamma}
\end{equation}
For a symmetrical APB,  the integral in (\ref{sigma}) or (\ref{Gamma}) can be written as
twice of the integral over positive $\eta$.

Relations similar to  (\ref{first-int})-(\ref{Gamma}) can also be derived for 
phases with several order parameters, such as the L1$_2$ or L1$_0$ phase. In particular, for
an APB separating two L1$_2$-ordered domains, the
 order parameters $(\eta_1,\eta_2,\eta_3)$ have the form
$(\zeta ,\eta ,\eta )$ mentioned above with the limiting values
$(\eta_0,\eta_0,\eta_0)$ and $(\eta_0,-\eta_0,-\eta_0)$ at $\xi\to\pm\infty$.
The variational equations and their first integral have the form 
analogous to  Eqs. (\ref{eqs_eta-c}) and (\ref{first-int}) but include three
functions, $c(\xi )$, $\zeta (\xi )$ and $\eta (\xi )$. It is again convenient 
to consider  $c$ and $\zeta$  as functions of  $\eta$ 
(while for  APBs in the L1$_0$ phase,  $c$ and $\eta$  as functions of  $\zeta$\,) 
and obtain a system of equations for  $c(\eta )$ and $\zeta (\eta )$ analogous to
COE  (\ref{COE}). Equations for $\xi (\eta )$, $\sigma$ and $\Gamma$  preserve their
form  (\ref{xi_eta})-(\ref{Gamma}) but $G (\eta )$ now
includes the derivative $\dot\zeta =d\zeta/d\eta$ and six functions
$g_{pq}$  which
are related to $g_{pq}^{\bot, a}$ in Eqs. (\ref{g-L1_2-MFA}),  (\ref{g-L1_2-PCA})
as follows:
\begin{equation}
g_{pq}(\alpha)=g_{pq}^{\bot}+g_{pq}^a\cos^2\alpha,
\label{g_pq-alpha}
\end{equation}
where $\alpha$ is the angle between the APB orientation and the local
tetragonality axis. When the nearest-neighbor interaction $v_1$ much exceeds the rest ones
(as in CuAu-based  alloys [9\ch 11]), the functions $g_{pq}(\alpha)$ are highly
anisotropic which is illustrated by Eqs. (\ref {g-L1_2-MFA}): $g_{\eta\eta}\sim \sin^2\alpha$; 
\ 
$g_{\zeta\zeta}\sim \cos^2\alpha$.
 It results in a notable anisotropy of distributions of
APBs, including the presence  of
many low-energy ``conservative'' APBs with  
 $\alpha\simeq 0$ in the  L1$_2$ phase and $\alpha\simeq \pi/2$
in the  L1$_0$ phase, as well as 
a peculiar alignment of APBs in ``twinned" L1$_0$ structures 
[9\ch 11]. In more detail, COE and
Eqs. (\ref{xi_eta})-(\ref{g_pq-alpha}) for the L1$_2$ and L1$_0$ phases will be  discussed
elsewhere.

Let us now discuss some applications  of  Eqs. (\ref{COE})-(\ref{g_pq-alpha}). 
First I consider the case
 when the equilibrium order parameter $\eta_0$ is 
small. Then the function  $f$ in (\ref{F_GL}) can be written as the Landau expansion:
\begin{equation}
f(\eta,c)=\varphi+a\eta^2+b\eta^4+d\eta^6
\label{f_eta}
\end{equation}
where $\varphi$, $a$, $b$ and  $d$ are some functions of concentration $c$ and temperature $T$. 
The equilibrium value $\eta_0$ 
is determined by the equation $f_{\eta}^0=0$, while
the ordering spinodal $T=T_s(c)$ (the disordered phase stability limit)
is  determined by the equation: $a(c,T)=0$.
Small values $\eta_0$ under consideration 
correspond to $c_0,T$ points  near the ordering 
spinodal where $a(c_0,T)$ is small. 

It is clear from both physical considerations and the results below that  the difference
$(c-c_0)$ at small $\eta_0$ is small, too. Therefore, functions
$(f_c-\mu)$ and  $f_{\eta}$ in Eq. (\ref{COE}) 
can be expanded in powers of $(c-c_0)$, $\eta$  and $\eta_0$:
\begin{eqnarray}
f_c-\mu&=&a_c^0\,(\eta^2-\eta_0^2)+\varphi_{cc}^0\,(c-c_0)+\ldots\\ 
f_{\eta}&=&4\eta\, b_0\,(\eta^2-\eta_0^2)+2\eta\, a_c^0\,(c-c_0)+\ldots
\label{f_c-eta}
\end{eqnarray}
Here $\varphi_{cc}^0$ is  $(\partial^2\varphi/\partial c^2)_0$,
and  dots mean terms of higher orders in $\eta_0^2$. 
The analogous expansion of  $\Omega (c,\eta)$ starts with terms bilinear in
$(c-c_0)$ and $(\eta^2-\eta_0^2)$, while for functions $g_{\eta c}$ and $\dot c$ 
the expansions start with terms linear in $\eta$. Thus terms with $\Omega$
and $f_{\eta}$ in (\ref{COE}) are proportional to $\eta_0^4$ being small
compared to  $(f_c-\mu)\sim \eta_0^2$, and COE is reduced to the equation
$f_c=\mu$,   which yields:
\begin{equation}
c_0-c(\eta )=(\eta_0^2-\eta^2)(-a_c^0)/\varphi_{cc}^0.
\label{c_eta}
\end{equation}
Taking  derivative of equation $a(c,T)=0$ one obtains:
 $(-a_c^0)=T_s^{\prime}\alpha$ 
where $\alpha$ is $(\partial a/\partial T)_0$ and  $T_s^{\prime}=dT_s/dc$. 
Thus the surface segregation at APB is proportional to the
ordering spinodal slope  $T_s^{\prime}(c_0)$, and so it
 decreases with approaching  the critical point where $T_s^{\prime}= 0$.

Let us now  suppose the alloy state $c_0,T$ in the $c,T$ plane 
to be close to the second-order transition line $T_s(c)$ 
 far from possible tricritical points.
Then  the higher-order terms in expansions  (\ref{f_eta})-(\ref{f_c-eta})
can be neglected, and for the function $\Omega (\eta)$ (\ref{Omega})  such expansion 
yields:
\begin{equation}
\Omega (\eta)=\tilde b\,(\eta_0^2-\eta^2)^2;\qquad \tilde b=b_0-(\alpha\,T_s^{\prime})^2/2\varphi_{cc}^0.
\label{Omega_eta}
\end{equation}
Using Eqs. (\ref{xi_eta})-(\ref{Gamma}) and (\ref{Omega_eta}) one obtains
 in this case for  $\eta (\xi)$, $c(\xi)$,
 the APB energy $\sigma$ and the segregation $\Gamma$:
\begin{eqnarray}
\eta (\xi)&=&\eta_0\tanh (\xi/\delta );\ \ 
c_0-c(\xi)=\eta_0^2\,\lambda\cosh^{-2} (\xi/\delta );
\nonumber\\
\sigma &=&\frac{8}{3v_a}\,\eta_0^3\left(g\,\tilde b\right)^{1/2};\ 
\Gamma =\frac{2}{v_a}\,\eta_0\,\lambda
(g/\tilde b)^{1/2}
\label{sigma-Gamma}
\end{eqnarray}
where $\lambda$ is $\alpha\,T_s^{\prime}/\varphi_{cc}^0$;
$g$ is $g_{\eta\eta}^0$; and 
$\delta =(g/\eta_0^2\tilde b)^{1/2}$ is the APB width.
These expressions generalize the earlier MFA results \cite{DMSV} to 
the case of any GL functional. The dependences $\eta (\xi)$ and $c (\xi)$ 
in Eqs. (\ref{sigma-Gamma}) are  similar to those observed in the Monte Carlo study
of segregation at APBs \cite{Schmid-Binder}. 
The temperature dependence of this segregation at small $\eta_0$ is more 
sharp than that of the APB energy: $\Gamma \propto\eta_0\sim (T_s-T)^{1/2}$
while  $\sigma\propto\eta_0^3\sim (T_s-T)^{3/2}$. Eqs. 
(\ref{Omega_eta})-(\ref{sigma-Gamma}) also show that the presence of
segregation results in a renormalization  of  the Landau
parameter $b_0$ entering  characteristics of APB to the lesser
value $\tilde b$ given by Eq. (\ref{Omega_eta}).
It results in a decrease of the APB energy $\sigma$ and an increase of its width
$\delta$ and segregation $\Gamma$ under decreasing temperature $T$ 
along the ordering spinodal $T=T_s(c)$.

The point $c_0,T$ at which both $a(c_0,T)$ in (\ref{f_eta}) and $\tilde b$ 
in (\ref{Omega_eta}) vanish corresponds to the tricritical point $c_t,T_t$.
At $T<T_t$ the second-order transition line $T_s(c)$ in the $c,T$ plane
splits into two binodals, $c_{bo}(T)$ and $c_{bd}(T)$, delimiting the single-phase 
ordered and disordered field, respectively. Such tricritical
point is observed, for example,  in  alloys Fe--Al
\cite{Allen-Cahn}. At this point the lowest
order terms in Eqs. (\ref{f_c-eta}) and (\ref{Omega_eta}) vanish,
one should consider the next-order terms, and function $\Omega$ (\ref{Omega})
at small $x=(c_0-c_t)$ and $t=(T-T_t)$ takes the form:
\begin{equation}
\Omega (\eta)=A\,(\eta_0^2-\eta^2)^2(\eta^2+h).
\label{Omega_eta-h}
\end{equation}
Here $h=h(x,t)$ is a linear function of $x$ and $t$ which can be written in terms of
the  binodal temperature derivative  $c_{bo}^{\prime}=dc_{bo}/dT$ as:
$h=\nu\,(x-t\,c_{bo}^{\prime})$,
while $A$ and $\nu$ are some positive constants.
Using Eqs. (\ref{xi_eta})-(\ref{Gamma}) and (\ref{Omega_eta-h})
one obtains for the characteristics of APB near $T_t$:
\begin{eqnarray}
\eta (\xi)&=&\frac{\eta_0\,\sinh y}{(\cosh^2y+\alpha)^{1/2}};\ \ 
c_0-c(\xi )=\frac{ \lambda\,\eta_0^2(1+\alpha)}{(\cosh^2y+\alpha)};\nonumber\\
%\label{eta-c_xi-t}
\sigma &=&\frac{J(\alpha )}{v_a}(gA\eta_0^4)^{1/2}; \ \ 
\Gamma =\frac{2\lambda L(\alpha)}{v_a}\,(g/A)^{1/2}.
\label{sigma-Gamma-t}
\end{eqnarray}
Here $y$ is $\xi/\tilde\delta$; $\alpha$ is $\eta_0^2/h$; while  $\tilde\delta$, $L(\alpha )$ and 
$J(\alpha )$ are:
\begin{eqnarray}
\tilde\delta=[g/A\eta_0^2(h&+&\eta_0^2)]^{1/2};
L(\alpha )=\ln\, [(1+\alpha )^{1/2}+\alpha^{1/2}];\nonumber\\
J(\alpha )=\frac{1}{2\alpha^2}[(1&+&4\alpha )L(\alpha )+(2\alpha -1)(\alpha +\alpha^2)^{1/2}].
\label{J_alpha}
\end{eqnarray}
Function $h(x,t)$ in  (\ref{Omega_eta-h})-(\ref{J_alpha}) is proportional
to the distance  in the $c,T$ plane from point $c_0,T$ to the binodal $c_{bo}(T)$,
while $\eta_0^2$ is proportional to the distance
to the ordering spinodal $T_s(c)$. Thus at small $\alpha\ll 1$, Eqs.
 (\ref{Omega_eta-h})-(\ref{J_alpha})  turn into 
 (\ref{Omega_eta})-(\ref{sigma-Gamma}) with $\tilde b=Ah$ and
describe the critical behaviour of APB near $T_s(c)$ discussed above.
The opposite case, $\alpha\gg 1$,
corresponds to the region of ``wetting"
APBs which received recently much 
attention \cite{Kikuchi-Cahn-79,Braun-Cahn,Le_Bouar}. 
Value $h=0$ corresponds to the ordered state with $c_0=c_{bo}(T)$,
and then COE (\ref{COE})   describes an IPB betwen this state and
the disordered state with $\eta_d =0$ and $c_d=c_{bd}(T)$. Substituting Eq. 
(\ref{Omega_eta-h}) with $h=0$ into 
Eqs. (\ref{xi_eta}) and (\ref{sigma}) one obtains for  this  IPB:
%
%\begin{equation}
\begin{eqnarray}
\eta (\xi)=\frac{\eta_0}{(1+e^{-z})^{1/2}}; \ \ \ 
c(\xi)&-&c_d=\frac{\lambda\,\eta_0^2}{(1+e^{-z})};\\
\sigma_{\rm APB}&=&2\sigma_{\rm IPB}.
\label{wet-rel}
%\end{equation}
\end{eqnarray}
Here $z$ is $\xi/\delta_1$;  $\delta_1=(g/A)^{1/2}/2\eta_0^2$ is the IPB width;
$\sigma_{\rm IPB}=(gA\eta_0^4)^{1/2}/2v_a$ is the IPB energy;
$c_d$ is $c_0-\lambda\eta_0^2$; and the coordinate
$\xi_1=0$ in (\ref{xi_eta}) is chosen at $\eta_1=\eta_0/\sqrt{2}$ 
where $c(\eta_1)$ is $(c_0+c_d)/2$. Eqs. (29)
show that  the order parameter $\eta$ in the disordered phase
decreases with moving from IPB  much
more slowly than the concentration deviation: $\eta\sim (c-c_d)^{1/2}$.
Eqs. (\ref{sigma-Gamma-t})-(\ref{wet-rel}) also show that in the ``wetting"
regime of large $\alpha$, the profiles $\eta (\xi)$ and $c(\xi)$ in Eq. 
(\ref{sigma-Gamma-t}) correspond to  the presence 
at $\xi_{\pm}=\pm\delta_1\ln\alpha$ of two almost independent IPBs described
by Eqs. (29). The total  width $l\simeq (\xi_+-\xi_-)$ and the segregation
$\Gamma$ for such APB are proportional to $\ln (1/h)$ while the energy difference 
$(\sigma_{\rm APB}-2\sigma_{\rm IPB})$ is proportional to $h\ln (1/h)$, 
which are usual dependences for 
the wetting regime \cite{Widom}. 
Eqs. (\ref{sigma-Gamma-t})-(\ref{wet-rel}) specify these relations for
the vicinity of tricritical points and enable one to follow the 
transition from wetting to the critical behaviour of APBs
under variation of  $T$ or $c_0$.

Let us now apply Eqs. (\ref{COE})-(\ref{sigma}) 
to derive the  wetting relation (\ref{wet-rel}) for the phases with several order parameters,
in particular, for the L1$_2$ phase in equilibrium with the L1$_0$ or the disordered FCC (A1) 
 phase.
 This problem was discussed by a number of authors
 \cite{Kikuchi-Cahn-79,Braun-Cahn,Le_Bouar}  but the general proof
seems to be absent yet. Let us first note that in consideration of IPB,
the initial condition to COE (\ref{COE})  can be put
in  either the ordered or the disordered phase, i.e. at $(c,\eta)$ values equal to either
$(c_0,\eta_0)$ or $(c_d,0)$, while the solution 
$c_{\rm IPB}(\eta)$ at either choice is the same and unique. Therefore,
in consideration of APB with the same initial values $c_0,\eta_0$,
the solution $c_{\rm APB}(\eta)$ coincides with $c_{\rm IPB}(\eta)$ at $\eta >0$,
it is $c_{\rm APB}(-\eta)$ at $\eta <0$, and so Eq. (\ref{wet-rel}) 
follows from Eq. (\ref{sigma}).
For the APB or IPB in the L1$_2$ phase, the local order can be described 
by the parameters $(c,\zeta,\eta)$ mentioned above, and
their initial values in COE are $(c_0,\eta_0,\eta_0)$, while the final ones
are $(c_0,\eta_0,-\eta_0)$, $(c_d,0,0)$, and $(c_{l},\eta_l,0)$ for the case of
an APB, IPB(L1$_2$-A1), and IPB(L1$_2$-L1$_0$), respectively, where 
$c_{l}$ and $\eta_l$ correspond to the second  binodal L1$_2$-L1$_0$.
 Therefore, the wetting relation
(\ref{wet-rel}) for the L1$_2$-A1 or L1$_2$-L1$_0$ phase equilibrium follows
from COE and Eq. (\ref{sigma}), just as for the single-order-parameter case.
Note, however, that at the given orientation ${\bf n}_0$
there are three types of  an APB in  the L1$_2$ phase
with the local  order $(\eta_1,\eta_2,\eta_3)$ of the form $(\zeta,\eta,\eta)$,
$(\eta,\zeta,\eta)$ or $(\eta,\eta,\zeta)$, and the structure and energy 
 for each  type is, generally, different. Therefore, there
are at least three types of an IPB(L1$_2$-A1)
 corresponding to ``a half" of the relevant APB in the wetting limit.
In the course of the kinetical wetting (for example, under
A1$\to$A1+L1$_2$ transformations studied in  \cite{Wang,BDPSV}) each APB first 
transforms into two ``its own" IPBs, but later on  these IPBs can evolve
to other types.

The effects of anisotropy under  wetting APBs
in Eqs. (\ref{xi_eta})-(\ref{Gamma}) are described by a factor $G^{1/2}$, while
the main contribution to  $\Omega (\eta)$ here is determined by the thermodynamic
relations. In particular,  singular
contributions to $\sigma$ and $\Gamma$ 
under wetting L1$_2$-APB by the A1 or L1$_0$  phase correspond to the region of small 
$\eta$ where $\Omega$ has the same form as in Eq. (\ref{Omega_eta-h}),
 while $G(\eta)$ is reduced to its first
term $g_{\eta\eta}$ as functions $g_{\eta\zeta}$, $g_{\eta c}$, $\dot\zeta$ and $\dot c$,
being odd in $\eta$,  vanish at small $\eta$. Thus the main
contributions to the APB width and energy take the form:
\begin{equation}
l\sim  g_{\eta\eta}^{1/2}\,\ln\, (1/h);\,\ 
(\sigma_{\rm APB}-2\sigma_{\rm IPB})\sim  g_{\eta\eta}^{1/2}\,h\ln\, (1/h),
\nonumber
%\label{l_alpha}
\end{equation}
where the angular dependence $g_{\eta\eta}$ is given by 
(\ref{g_pq-alpha}). Therefore, the wetting effects can reveale a significant
anisotropy, particularly in the short-range interaction systems.
It  agrees with some previous results \cite{Kikuchi-Cahn-79,Braun-Cahn,Le_Bouar}.

The author is much indebted to I.R. Pankratov
for the  help in this work.
The work was supported  by the Russian Fund of Basic Research under
Grants No. 00-02-17692 and 00-15-96709.

\end{document}